\documentclass{PoS}
\newcommand{\be}{\begin{equation}}  
\newcommand{\ee}{\end{equation}}  
\newcommand{\beq}{\begin{eqnarray}}  
\newcommand{\eeq}{\end{eqnarray}}
\newcommand{\Dlr}{\buildrel \leftrightarrow \over D\raise-1pt\hbox{}}

\usepackage{amsmath}
\usepackage{amsfonts,amssymb}
\title{Sigma-terms and axial charges for hyperons and charmed baryons}

\ShortTitle{Sigma-terms and axial charge for hyperons and charmed baryons}

\author{Constantia  Alexandrou \\
         Department of Physics, University of Cyprus, P.O. Box 20537, 1678 Nicosia, Cyprus, and\\  
 Computation-based Science and Technology Research  
    Center, Cyprus Institute, 20 Kavafi Str., Nicosia 2121, Cyprus \\  
        E-mail: \email{alexand@ucy.ac.cy}}

\author{\speaker{Kyriakos Hadjiyiannakou}\\
       Department of Physics, University of Cyprus, P.O. Box 20537, 1678 Nicosia, Cyprus\\
        E-mail: \email{hadjigiannakou.kyriakos@ucy.ac.cy}}
\author{Karl Jansen\\
       NIC, DESY, Platanenallee 6, D-15738 Zeuthen, Germany\\
        E-mail: \email{karl.jansen@desy.de}}
\author{Christos Kallidonis\\
        Department of Physics, University of Cyprus, P.O. Box 20537, 1678 Nicosia, Cyprus, and\\  
 Computation-based Science and Technology Research  
    Center, Cyprus Institute, 20 Kavafi Str., Nicosia 2121, Cyprus\\
        E-mail: \email{kallidonis.christos@ucy.ac.cy}}

\abstract{We present results for the $\sigma$-terms and axial charges  for various hyperons and charmed baryons using $N_f=2+1+1$ twisted mass fermions.
For the computation of the three-point function we use the fixed current
 method. For one of the $N_f=2+1+1$ ensembles with pion mass of 373~MeV we compare the results
of the fixed current method with those obtained with 
 a stochastic method for computing the all-to-all propagator involved in
the evaluation of the  three point functions.}

\FullConference{31st International Symposium on Lattice Field Theory LATTICE 2013\\
                 July 29 - August 3, 2013\\
                 Mainz, Germany}

\begin{document}
\section{Introduction}
The nucleon axial charge $g_A$ is a well-measured quantity extracted from neutron $\beta$- decay experiments. The nucleon light $\sigma_{\pi N}$ term 
is extracted from $\pi N$ scattering phase shifts combining experimental  measurements and
phenomenology. The nucleon strange content $\sigma^s$ is
 extracted in a similar manner using the kaon $N$ scattering phase shifts, which however are
difficult to measure and, furthermore, 
chiral perturbation theory may not be applicable. 
For other particles, these quantities are either poorly known or their values 
are not known experimentally. Assuming SU(3) flavor symmetry  one can obtain useful relations among the axial charges of hyperons. Such relations are used as input in low-energy effective theories~\cite{Tiburzi:2008bk, Alexandrou:2009qu} and it is therefore important to test the degree
of their validity.
 Lattice QCD provides the appropriate framework to calculate these
quantities for all baryons. In this work, we thus focus on computing
the $\sigma$ -terms and axial charges for hyperons and charmed baryons.

While the evaluation of the nucleon axial charge $g_A$ has been carried out  by a number of lattice QCD collaborations~\cite{Alexandrou:2010cm} 
 the axial charges of other baryons are not
so well studied. The focus of this work is to
evaluate the axial charges and $\sigma$-terms, which are extracted from matrix
elements at zero momentum transfer. Since we are interested in
evaluating matrix elements for any baryon, the  fixed current method
is the appropriate approach yielding  with one sequential inversion per quark flavor
the axial charges of all baryons. An additional sequential inversion per quark flavor
is carried out to extract the $\sigma$-terms. In order 
to avoid additional sequential  inversions are for every operator  we  study a new method for computing hadron 
matrix elements using a stochastic method to calculate the all-to-all propagator
entering the three-point function.  We will refer to this
 alternative method as the stochastic method, which  is extremely versatile as compared to either the fixed sink or fixed current approaches conventionally employed
in three-point function computations. The advantage of the stochastic method
is that once the all-to-all propagator is computed one can obtain the matrix elements of any hadron and for any operator.

\section{Tuning the strange and charm quark mass}
\begin{figure}[h!]
\begin{minipage}{0.49\linewidth}
 \includegraphics[width=\linewidth]{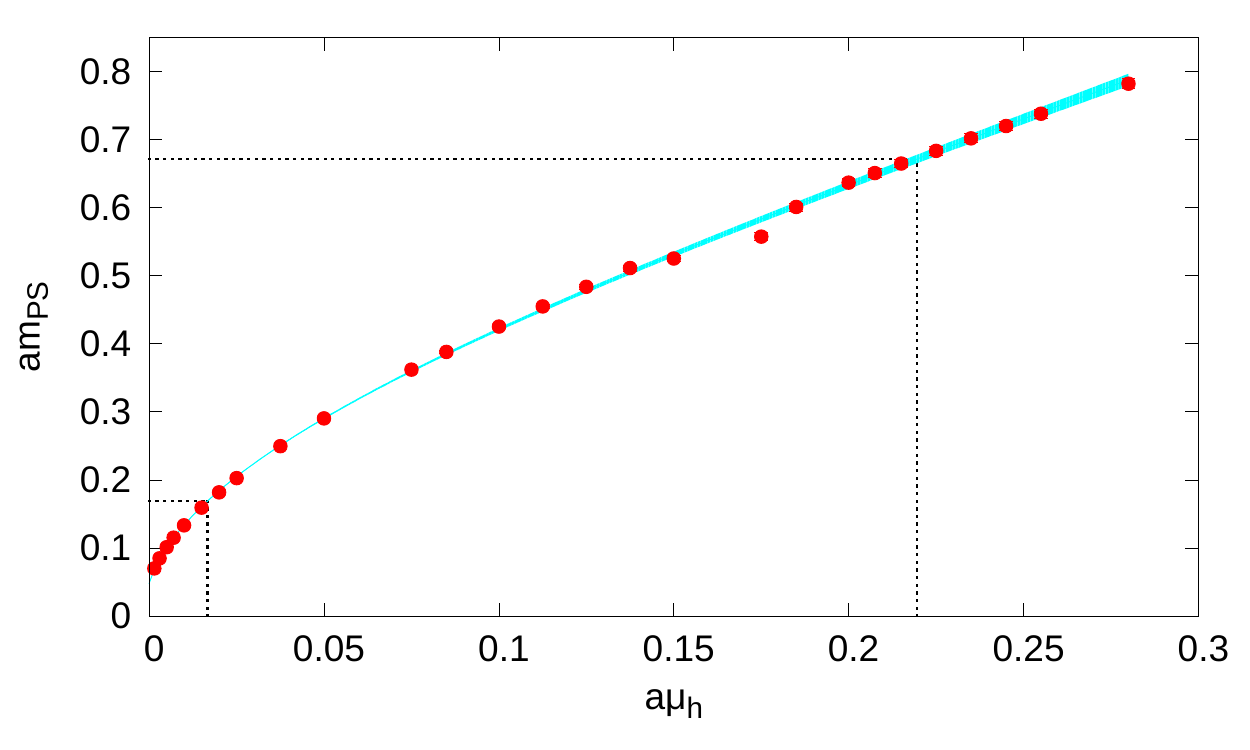}
\end{minipage}
\begin{minipage}{0.49\linewidth}
 {\includegraphics[width=\linewidth]{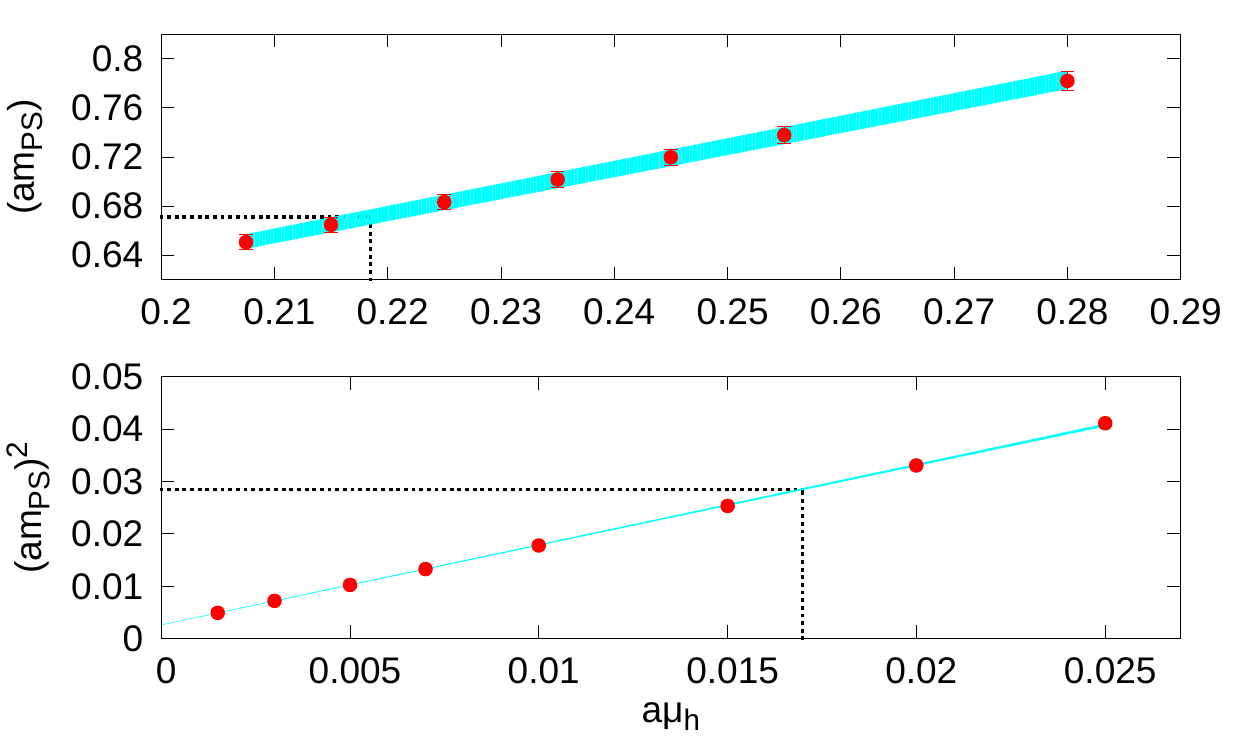}}  
\end{minipage}
 \caption{Pseudoscalar mass as a function of $a\mu_h$. Left: Fit to the  polynomial given in Eq.(2.2). Right: Linear fit around the range of the kaon mass (upper) and  around the D-meson mass (lower).}
\label{fig:tuning}
\end{figure}

We use $N_f=2+1+1$ twisted mass fermions (TMF) configurations simulated at
$\beta=1.95$. The action we employ introduces a twisted heavy mass-split doublet for the strange and the charm quark
\begin{equation}
S_F^{(h)} [ \chi^{(h)},\bar{\chi}^{(h)},U] = a^4 \sum_x \bar{\chi}^{(h)}(x) \left( D_W[U] + m_0 + i\mu_\sigma \gamma_5 \tau^1 + \tau^3 \mu_\delta \right) \chi^{(h)}(x)\>,
\end{equation}
where $m_0$ is the bare untwisted quark mass, $\mu_\sigma$ is the bare twisted mass along the $\tau^1$ direction and $\mu_\delta$ is the mass splitting in the $\tau^3$ direction~\cite{Baron:2010bv}. For the valance sector we use Osterwalder-Seiler fermions and, therefore, we need to tune  the strange and charm mass.
 The tuning is performed by matching the mass of the kaon and D-meson in  the
 tuunitary and mixed action setups. To perform the matching we vary the mass of
the strange and charm quarks and fit the resulting masses  
using the following polynomial form~\cite{Blossier:2007vv}
\begin{equation}
a^2 M_{PS}^2 (a\mu_l,a\mu_h) = a_1(\mu_l+\mu_h) + a_2(\mu_l+\mu_h)^2 + a_3(\mu_l+\mu_h)^3 + a_4(\mu_l+\mu_h)(\mu_l-\mu_h)^2\, ,
\label{polynomial}
\end{equation}
where $\mu_l$ is the mass of the u- and d- quarks as used in the
simulation and $\mu_h$ is varied in a range that includes the kaon and D-meson mass. The resulting fit can be seen in Fig.~\ref{fig:tuning} covering the whole range of
masses from the kaon to the D-meson mass.  As a consistency check we also perform a fit to a smaller range in $a\mu_h$
close to the kaon mass using $a^2M_{PS}^2(a\mu_s) = c_1 + c_2 a\mu_s$  and independently around the D-meson mass using $aM_{PS}(a\mu_c)=d_1+d_2 a\mu_c$ as shown
in the right panel of Fig.~\ref{fig:tuning}. We find $a\mu_s=0.01671(18)(28)$
and $a\mu_c=0.2195(15)(10)$, where the first error is statistical and the second
systematic determined as the difference between the value extracted from the polynomial fit over the whole range of heavy masses and the one extracted from the linear fit over the range shown in the right panel in Fig.~\ref{fig:tuning}.




\section{Axial charges}
\begin{figure}[h!]
\begin{minipage}{0.49\linewidth}
\includegraphics[width=\linewidth]{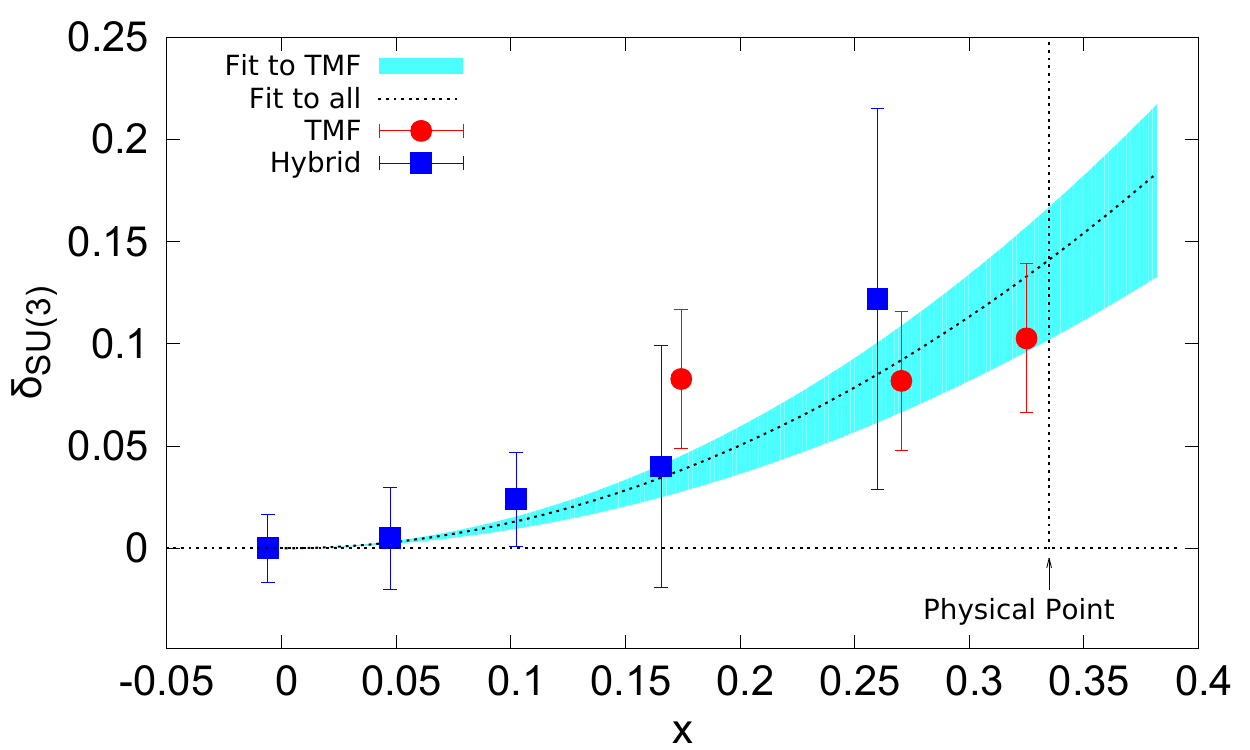}
\end{minipage}\hfill
\begin{minipage}{0.49\linewidth}
\includegraphics[width=\linewidth,height=0.6\linewidth]{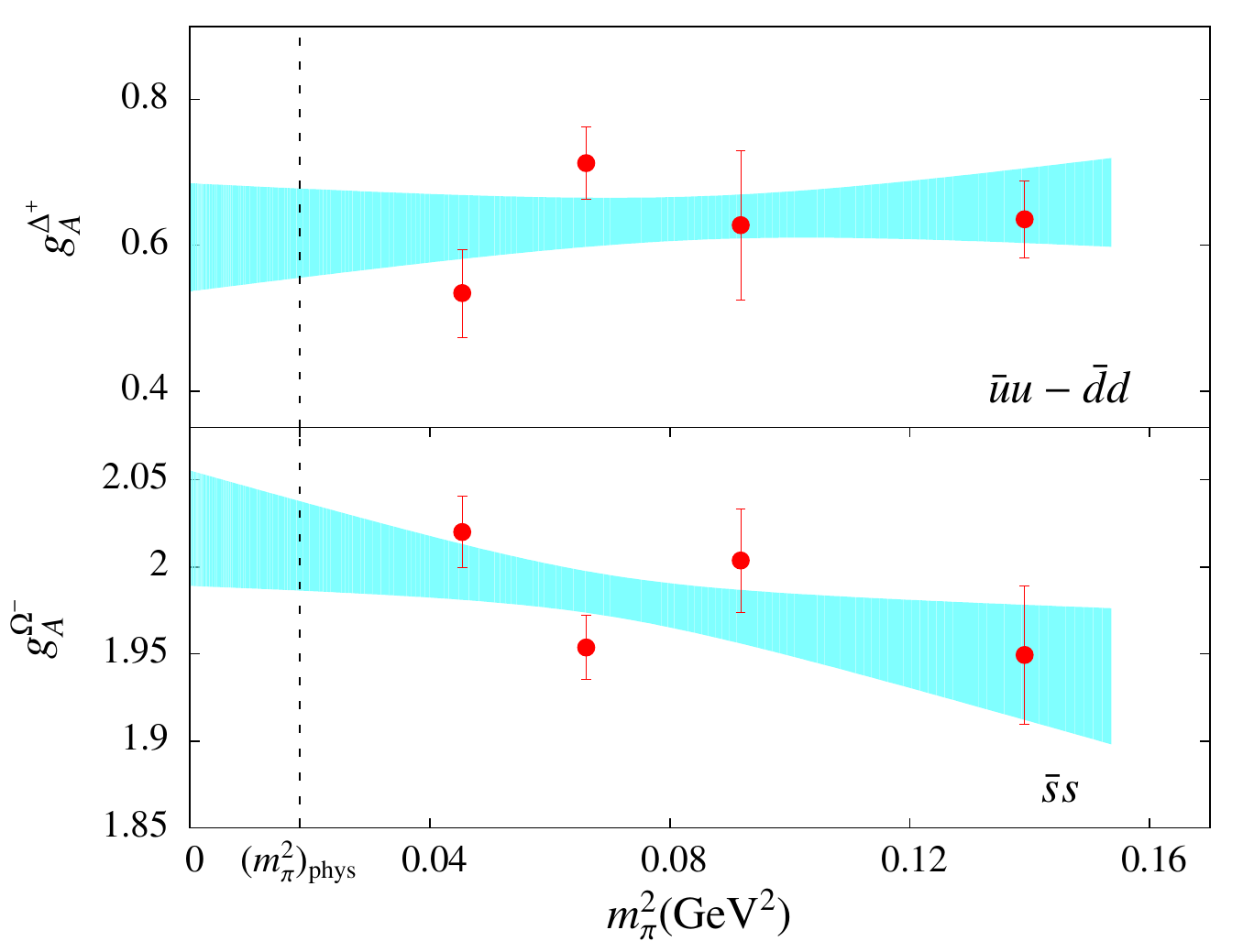}
\end{minipage}
\caption{Left: $\delta_{\rm SU(3)}$ versus $x=(m_{K}^2-m_{\pi}^2)/(4\pi^2 f_\pi^2)$. Filled red circle are TMF results and filled blue squares are results
using a hybrid action~\cite{Lin:2007ap}. Right: Isovector axial charge of the $\Delta^+$ and the connected contribution to the strange 
axial charge of the $\Omega^-$  versus $m_\pi^2$.
}
\label{fig:SU(3) symmetry}
\end{figure}

The axial charge $g_A$ is a fundamental quantity of the structure of a hadron.
 Within lattice QCD
the  calculate the axial charge
 for any hadron, 
  is extracted from 
the matrix element  of the axial-vector current, 
$A^3_\mu \equiv \bar{\psi}(x) \gamma_\mu \gamma_5  \psi(x)$ at zero momentum
transfer: $\langle h(\vec{p}^\prime)|A_\mu|h(\vec{p})\rangle$. $SU(4)$ flavor symmetry leads to two 20-plets  of spin-1/2 and spin-3/2 baryons, for which
we calculate the axial charges using the interpolating field given in Ref.~\cite{Alexandrou:2009qu} including a spin-3/2 projection, which is
found to be important in the case of the $\Xi^*$s.
Assuming SU(3) flavor symmetry  the axial charges of the low-lying octet baryons obey the following relations~\cite{Tiburzi:2008bk}
\begin{equation}
g_A^N = F + D, \,\, g_A^\Sigma = 2F,\,\, g_A^\Xi = -D+F \,\, \Longrightarrow \,\, \delta_{\rm SU(3)}\equiv g_A^N - g_A^\Sigma + g_A^\Xi = 0.
\end{equation}
We examine how well SU(3) flavor symmetry is satisfied for unequal 
quark masses by computing the SU(3) symmetry breaking parameter $\delta_{\rm SU(3)}$. In Fig.~\ref{fig:SU(3) symmetry} we plot $\delta_{\rm SU(3)}$
 as a function of the dimensionless parameter $x=(m_{K}^2-m_{\pi}^2)/(4\pi^2 f_\pi^2)$. In the plot we include results
obtained using a hybrid action of staggered fermions and domain wall valence
quark that includes a calculation at the SU(3) flavor symmetric limit~\cite{Lin:2007ap}. Performing a chiral extrapolation using $\delta_{\rm SU(3)}=ax^2$
we  find that the deviation from SU(3) flavor symmetry at the physical
value of the strange quark mass  is around $10\%$.

One can also compute the axial charges
of baryons belonging to the decuplet. 
We show in Fig.~\ref{fig:SU(3) symmetry} two representative examples
of the light and strange quark contribution to the axial charge  of the $\Omega^-$ and
the $\Delta^+$ computing only the connected contributions. For $\Omega^-$ 
the value  increases as we approach the physical pion mass. A linear fit to $m_\pi^2$ yields a good fit to the data and
provides a prediction of the strange axial charge of these baryons.  

Using the same techniques we compute the axial charges of charmed baryons 
whose values are not known. The interpolating fields are given in Ref.~\cite{Alexandrou:2012xk}.

\begin{figure}[h!]
\begin{minipage}{0.49\linewidth}
\hspace*{-0.5cm}\includegraphics[width=1.05\linewidth]{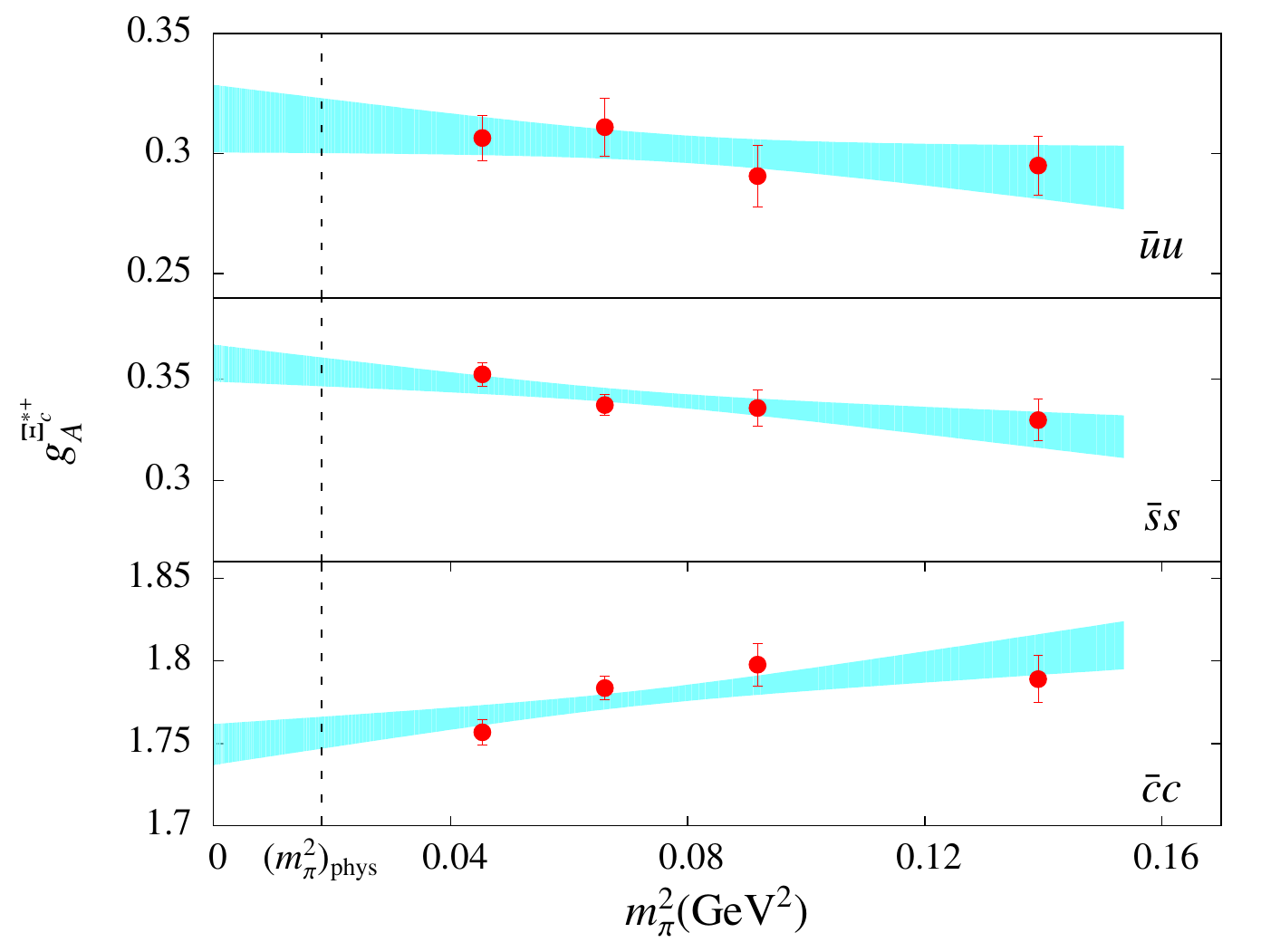}
\end{minipage}\hfill
\begin{minipage}{0.49\linewidth}
\hspace*{-0.5cm}\includegraphics[width=1.05\linewidth]{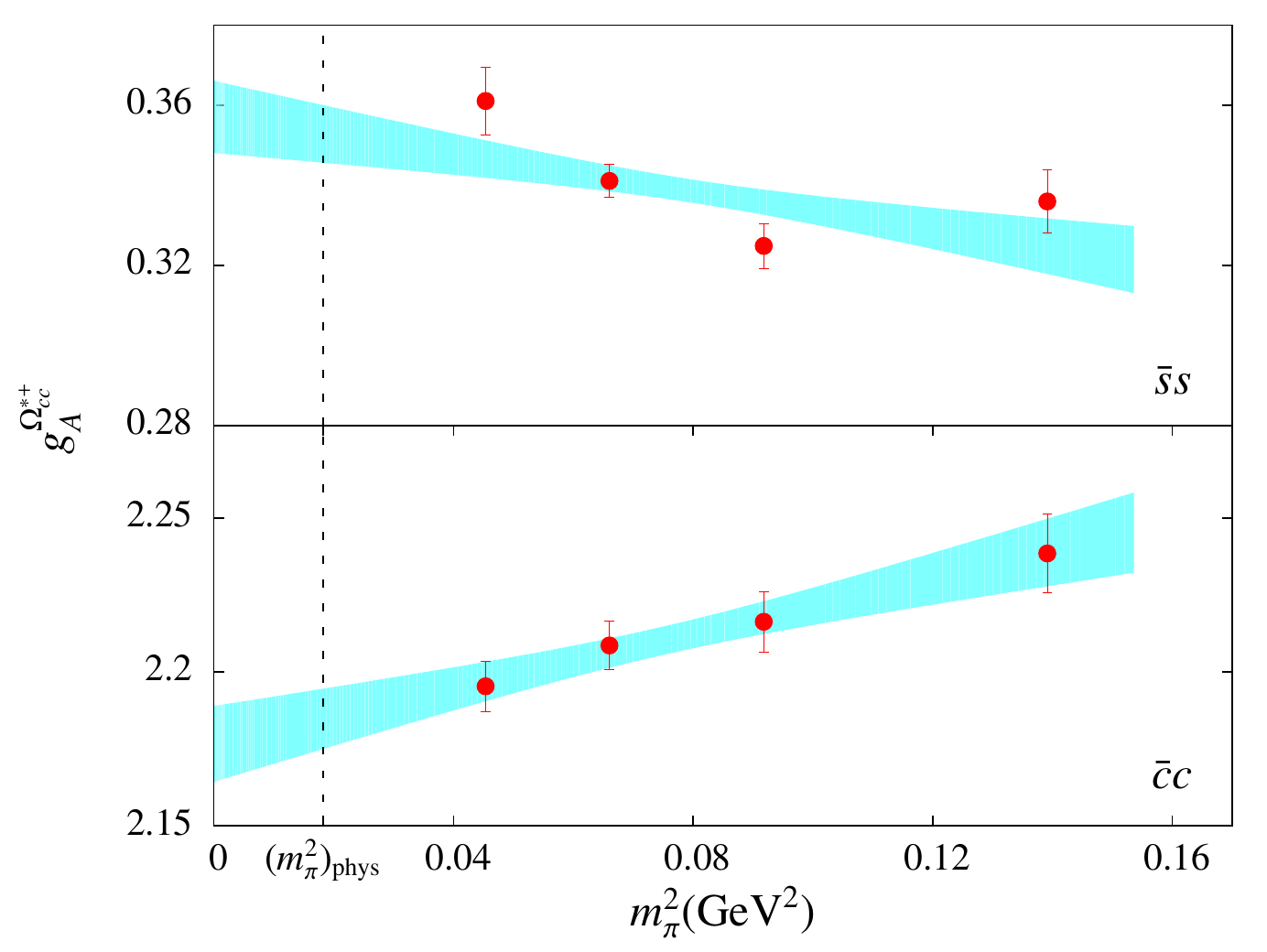}
\end{minipage}
  \caption{The axial charge as a function of $m_\pi^2$ for $\Xi^{*+}_c$ (left) and $\Omega^{*+}_{cc}$ (right).}
\label{fig:gA charmed}
\end{figure}

We show in Fig.~\ref{fig:gA charmed} two representative examples
of the strange and charm quark contribution to the axial charge of the $\Xi^{*+}_c$ and the $\Omega^{*+}_{cc}$ as a function of $m_\pi^2$. An interesting feature
is that the light and strange quark contributions to th e axial charge
of charmed baryons  increase as the pion mass approaches the physical value
while for the charm content decreases.

\section{$\sigma$-terms}
Experimental searches for cold dark matter need as input  the strength of the interaction between a WIMP and a nucleon mediated by a Higgs exchange.
Thus a reliable calculation of the nucleon $\sigma$-terms 
provides an  important input for these experiments. There are phenomenological determinations of the value of  $\sigma_{\pi N}$, as well as, lattice
 QCD calculations
mainly using the Feynman-Hellman theorem~\cite{Young:2013nn}, but also
 by directly calculating the nucleon matrix element $\sigma_{\pi N} = m_l \langle N\vert \bar{u}u + \bar{d}d \vert N \rangle$~\cite{Alexandrou:2012gz,Alexandrou:2013nda}.
In this work, we compute the 
 matrix element $\sigma_q=m_q\langle h \vert \bar{\psi} \psi \vert h \rangle$ for all low-lying baryons belonging to the two 20-plets of SU(4).

\begin{figure}[h!]
\begin{minipage}{0.49\linewidth}
\hspace*{-0.5cm}\includegraphics[width=1.2\linewidth]{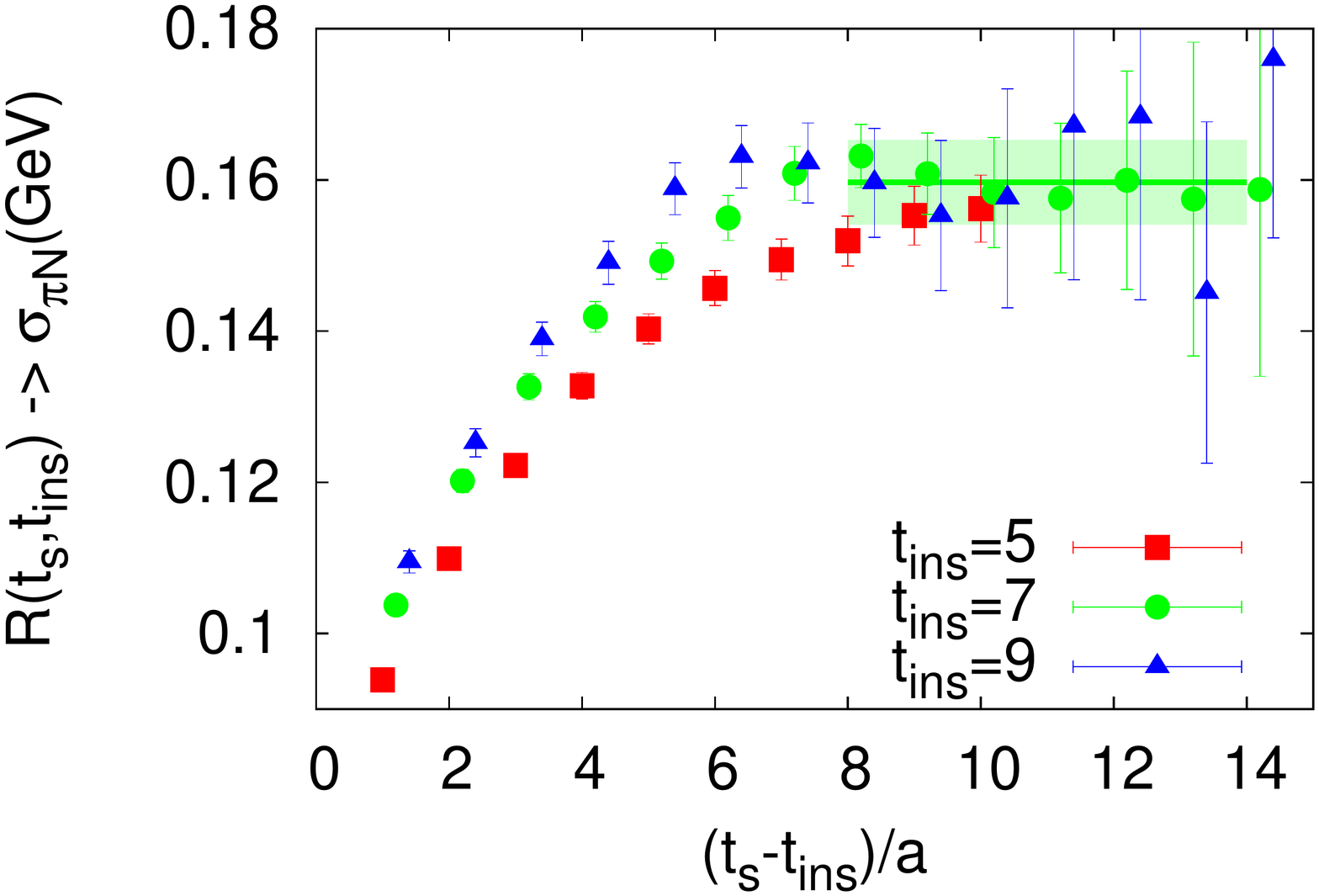}
\end{minipage}\hfill
\begin{minipage}{0.49\linewidth}
\hspace*{-0.5cm}\includegraphics[width=1.2\linewidth]{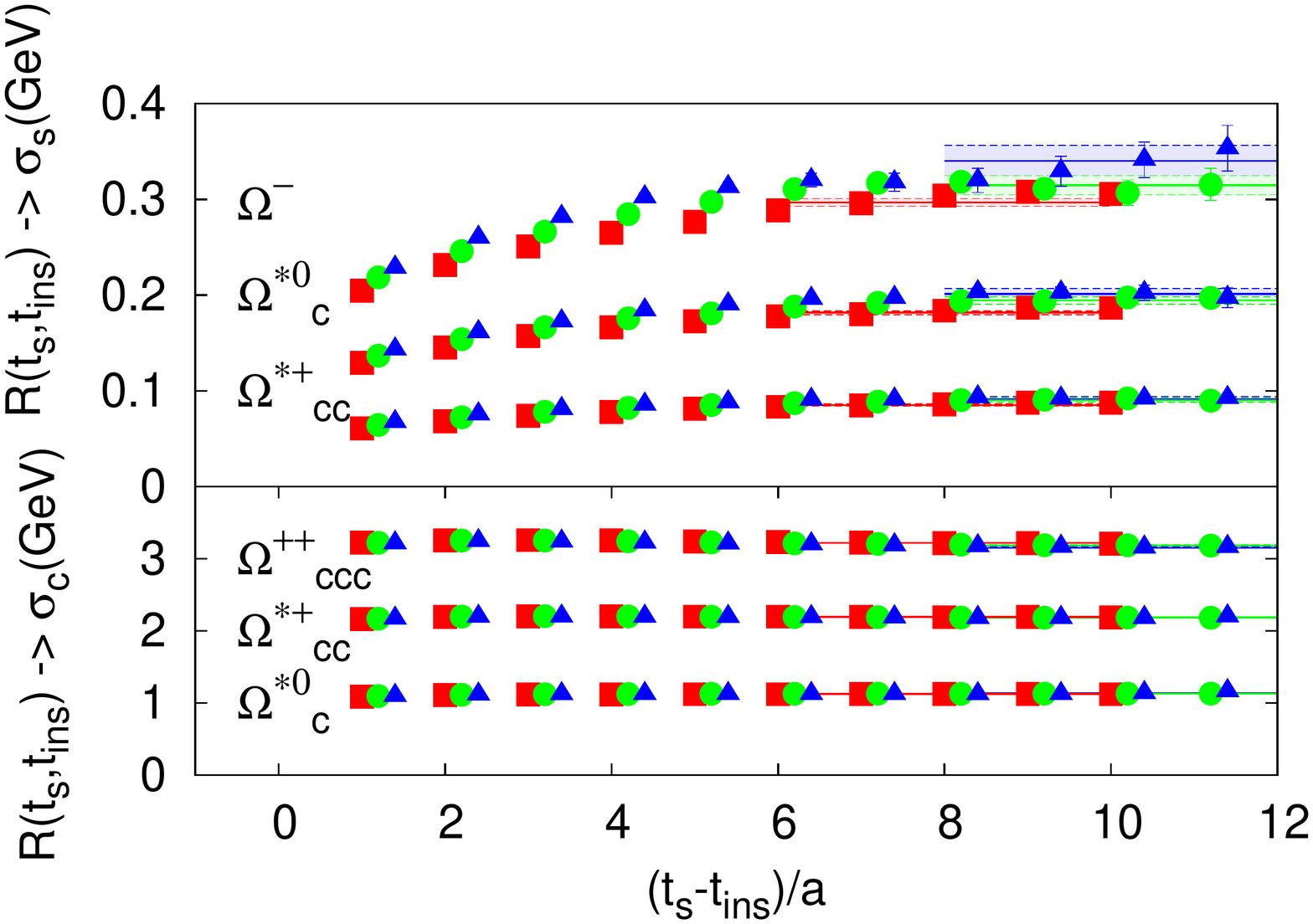}
\end{minipage}
  \caption{The ratio $R(t_{ins},t_s)$ which yields the $\sigma$-term as a function of $(t_s-t_{ins})/a$ for the nucleon (left) and for the strange (upper) and charm (lower) content of the  $\Omega$s (right).}
  \label{fig:sigmas}
\end{figure}

In Fig.~\ref{fig:sigmas} we show  the ratio of the three-point function to the two point function  $R(t_{ins},t_s)=\frac{G^{3pt}(\Gamma, t_{ins},t_s)}{G^{2pt}(t_s)}$ for representative cases for the light, strange and charm quark content, considering only connected contributions. In the fixed current approach one needs to fix the
time separation between current insertion and source, taken at zero, $t_{ins}$. Since this observable, unlike the axial charge, receives large excited state contributions we
need to ensure that $t_{ins}$  is sufficiently large so that
the excited states are  damped out before we extract
the value of $\sigma$ by fitting to a constant. We show the ratio as a function
of $t_s-t_{ins}$ in Fig.~\ref{fig:sigmas} for three values of $t_{ins}$. As can be seen $t_{ins}=7a$ yields consistent results with those for $t_{ins}=9a$. 
This is true for $\sigma_{\pi n}$, as well as,  for the strange and charm content of the
$\Omega$s.
Thus, we fix $t_{ins}=7a$.
In Fig.~\ref{fig:sigmas} we show the strange quark contribution to the $\Omega^-$
 which has three strange quarks, $\Omega^{*0}_c$ with two strange quarks and $\Omega^{*+}_{cc}$ with one strange quark. As can be seen the
strange quark contribution triples for $\Omega^-$ as compared to $\Omega^{*+}_{cc}$ as expected. The same is true also for the case of the charm contribution
to the  $\sigma$-term. 

\section{Stochastic Method for connected diagrams}

\begin{figure}[htbp]
\centerline{\includegraphics[width=0.9\linewidth]{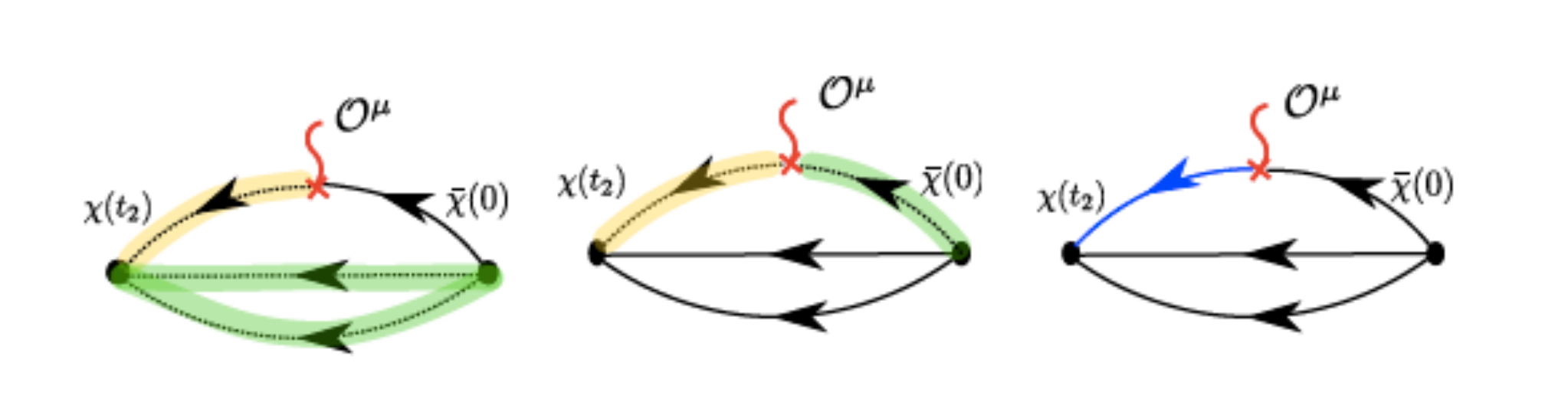}}
  \caption{Fixed sink (left), fixed insertion (center) and stochastic method (right). With green lines we depicted the part of the diagram which is used as a source for the sequential inversion  and with yellow the sequential propagator.}
\label{fig:diagrams}
\end{figure}

\begin{figure}[h!]
\begin{minipage}{0.49\linewidth}
\hspace*{-0.5cm}\includegraphics[width=1.2\linewidth]{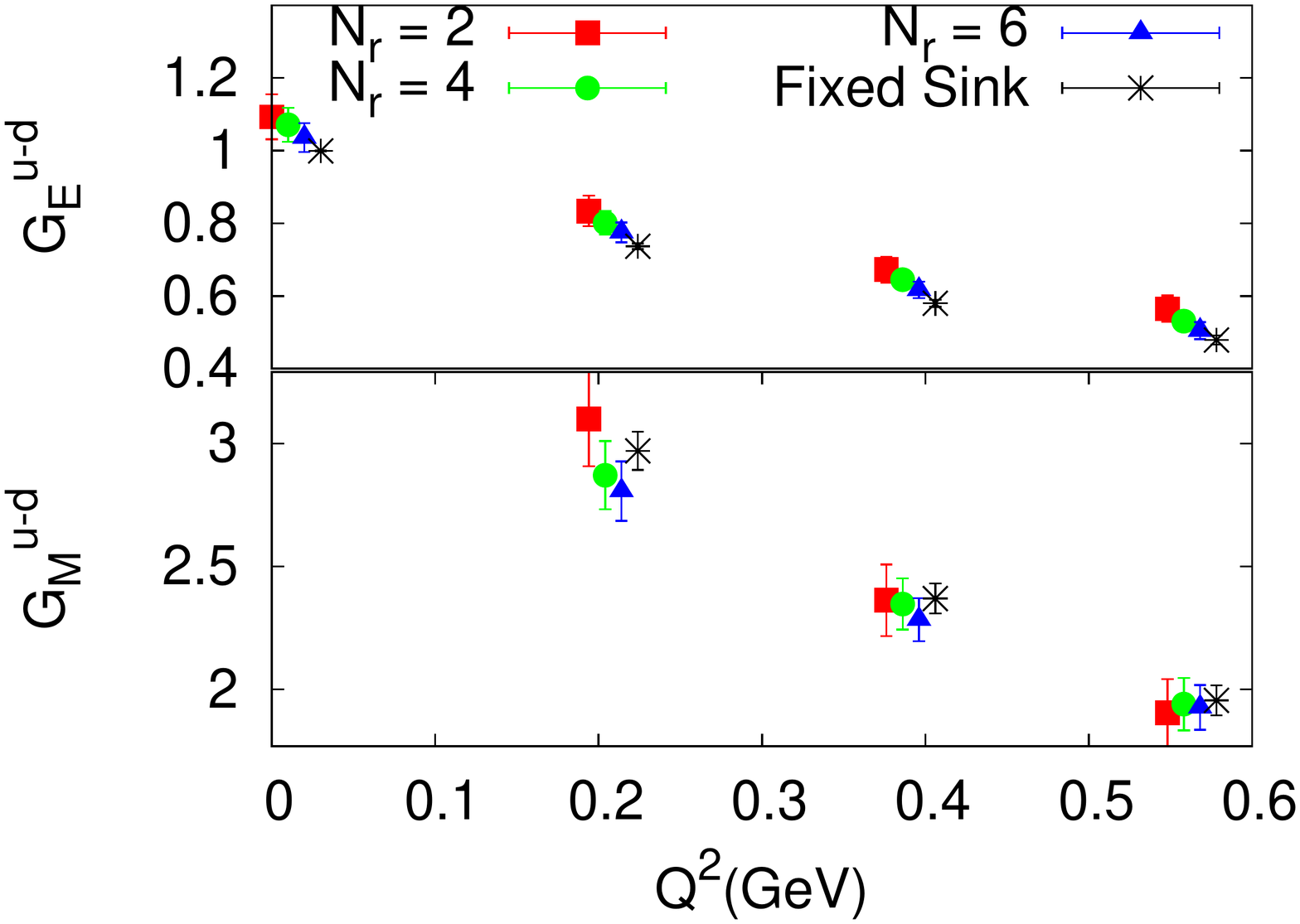}
\end{minipage}\hfill
\begin{minipage}{0.49\linewidth}
\hspace*{-0.5cm}\includegraphics[width=1.2\linewidth]{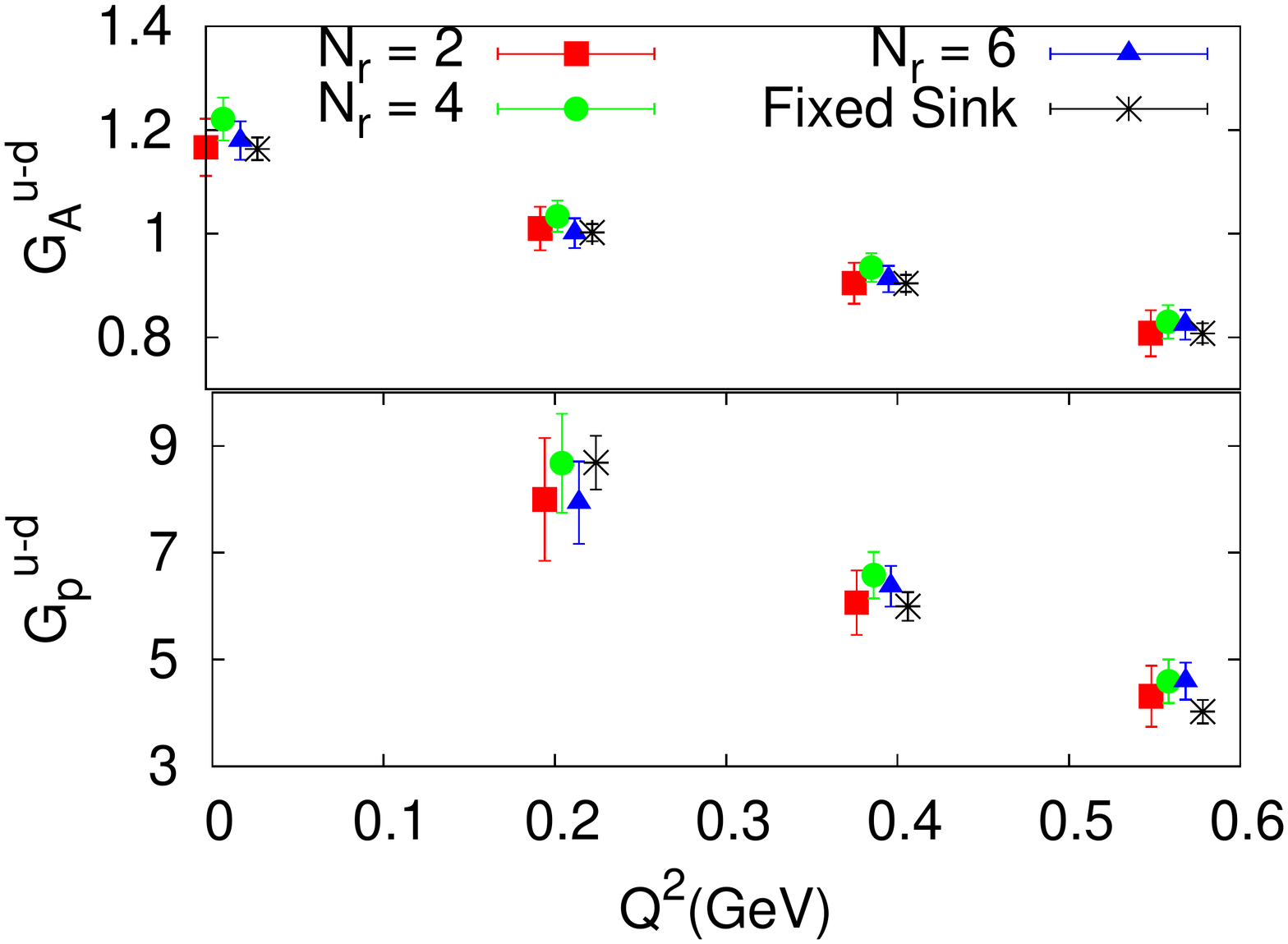}
\end{minipage}\vspace*{-0.3cm}
  \caption{Electromagnetic (left) and axial (right) form factors versus $Q^2$.}
  \label{fig:stochastic}\vspace*{-0.2cm}
\end{figure}

In Fig.~\ref{fig:diagrams} we show diagrammatically  the fixed sink and current methods that are used in the calculation of three-point functions.
The former requires us to fix the sink i.e. the hadron state
and the latter the operator insertion. In order to be able to compute
the three-point function for {\it every} hadron state {\it and current} insertion
we examine a third approach that computes the all-to-all stochastically~\cite{Alexandrou:2013xon}. Within this approach the all-to-all propagator
is written in terms of  a solution vector $\phi(x)$ and a noise vector $\xi(y)$
with the  consequence that the double sum involved in the calculation of three point functions becomes two single sums:

\begin{equation}
\sum_{\vec{y}} \sum_{\vec{x}} e^{-i\vec{p}^\prime \cdot \vec{x}} e^{-i\vec{p} \cdot \vec{y}} G(x;y)\Gamma G(y;0) \longrightarrow \\
 \frac{1}{N_r} \sum_{r=1}^{N_r} \left( \sum_{\vec{x}} e^{-i\vec{p}^\prime \cdot \vec{x}} \phi_r(x) \right) \left(    
 \sum_{\vec{y}} e^{-i\vec{p} \cdot \vec{y}} \xi^*_r(y) \Gamma G(y;0) \right).
\end{equation} 
This makes the calculation via the stochastic method feasible, provided the
number of noise vectors $N_r$ needed is small enough.
In Fig.~\ref{fig:stochastic} we compare results using the stochastic method to those obtained using the fixed sink method. The comparison is done using 500 gauge  configurations of $N_f=2+1+1$ TMF at pion mass $373$~MeV. We invert at every $t_s$ (time dilution) and consider  spin and color dilution. Thus for a given $t_s$  we need 12 inversions per noise vector which is the same
as the number of inversions need  to obtain the sequential propagator. 

In Fig.~\ref{fig:stochastic} we show results on the
nucleon electromagnetic and axial-vector form factors using $N_r=2,4$ and 6.  The results obtained using the stochastic  method  show good convergence obtaining 
 when  using $N_r=4$ and $N_r=6$ values that are consistent with those obtained with the fixed sink method for all cases
except the electric form factor, which requires
$N_r=6$. Taking $N_r=6$, means that the stochastic method needs six times more
 inversions to achieve the accuracy of the fixed sink method. However, the  stochastic method is more versatile and we can extract more measurements without additional inversions that over-compensate for this factor of six. To understand the
gain we consider the cost 
 needed to evaluate the axial form factors.


\begin{table}
\begin{center}\hspace*{-0.4cm}
	\begin{tabular}{|c|c|c|c|c|c|}
	\hline
	 & Projectors & States & \# $N_r$ & \# Inversions & Err(Stoch)/Err(Fixed Sink) \\ \hline
	 Fixed Sink & $\sum_{k=1}^3 \Gamma_k$ & $p$ & - & $24$ & - \\ \hline
	 Stochastic & $\sum_{k=1}^3 \Gamma_k$ & $p$ & $2$ & $2\times24$ & $\thicksim 3$ \\ \hline
	 Stochastic  & $\Gamma_1,\Gamma_2,\Gamma_3$ & $p,n$ & $2$ & $2\times24$ & $\thicksim 1$ \\ \hline
	\end{tabular}
\caption{The number of inversions and the resulting ratio of the error of the stochastic method to that of the fixed sink method for the summed and unsummed spin projectors $\Gamma_k$. }
\label{tab:comparison}\vspace*{-0.5cm}
\end{center}
\end{table}
In the case of the stochastic method to use three different projector $\Gamma_k$, $k=1,\cdots,3$ needs no extra inversions unlike the fixed-sink method. 
In Table~\ref{tab:comparison} we compare the results
obtained using the summed  projector $\sum_{k=1}^3\Gamma_k$  with the fixed sink method, which requires one inversion per quark flavor, with those obtained using
  the stochastic method. We find  about 3 times the error
  when using the stochastic method with $N_r=2$, which is twice as expensive as the fixed sink method. 
 However, in
 the stochastic method we can use the unsummed projector, which carries less noise than the summed one, and average over proton and neutron without extra cost  reducing  the error by a factor of about 3.  Thus, in the case the nucleon the cost is twice that of  the fixed sink 
technique for a fixed error. Thus, considering even only two hadron states we break even. Given that we can compute all form factors for the 40 SU(4) particles with no additional inversions shows the superiority of the stochastic method.

\section{Conclusions}
In this work we present results on the $\sigma$-terms and axial charges for
the two SU(4) 20-plet  baryons. We study the SU(3) flavor breaking and the behavior of the $\sigma$-terms as we vary the number of quarks.
For the extraction of the axial charges and $\sigma$-terms we use the fixed current method. To avoid the limitations of the fixed current or sink methods
which additional inversions for each operator or hadron state (spin projection) respectively we tested a stochastic method to calculate the all-to-all propagator involved in the computation of
connected three-point function. This  method  shows a very fast convergence to the results of the fixed sink method, and for the case of the nucleon axial charge, it is only twice
as expensive.  Therefore, it can be
considered as  an alternative approach for the calculation of three-point functions,
in particular when we are interested in matrix elements of more than one hadron.

\end{document}